\documentclass[conference]{IEEEtran}

\usepackage{graphicx}
\usepackage{subcaption}
\usepackage{fixltx2e}
\usepackage{hyperref}
\usepackage{balance}
\usepackage{colortbl}
\usepackage{tabularx}
\usepackage{tabu}
\usepackage{booktabs}% for better rules in the table
\usepackage{xparse}
\usepackage{multirow}
\usepackage{enumitem}
\usepackage[table,xcdraw]{xcolor}
\usepackage{xcolor}
\usepackage{fontawesome}
\usepackage{xpatch} 
\usepackage{blindtext}
\usepackage{graphicx}
\usepackage{calc}

\definecolor{pblue}{rgb}{0.13,0.13,1}
\definecolor{pgreen}{rgb}{0,0.5,0}
\definecolor{pred}{rgb}{0.9,0,0}
\definecolor{pgrey}{rgb}{0.46,0.45,0.48}
\definecolor{codebackground}{rgb}{0.95, 0.95, 0.92}
\definecolor{gray50}{gray}{.5}
\definecolor{gray40}{gray}{.6}
\definecolor{gray30}{gray}{.7}
\definecolor{gray20}{gray}{.8}
\definecolor{gray10}{gray}{.9}
\definecolor{gray05}{gray}{.95}
\definecolor{arsenic}{rgb}{0.23, 0.27, 0.29}

\newenvironment{rqbox}{\par\begingroup
\setbox0=\vbox\bgroup\noindent
\hsize=0.95\linewidth
\begin{minipage}{0.95\linewidth}\normalsize}
{\end{minipage}\egroup
\textcolor{gray20}{\fboxsep1.5pt\fbox
    {\fboxsep5pt\colorbox{gray05}{\normalcolor\box0}}}
\endgroup\par\noindent
\normalcolor\ignorespacesafterend}

\RequirePackage{fontawesome}
\newcommand{\ie}{i.e.,\xspace}
\newcommand{\eg}{e.g.,\xspace}

\newcommand{\etal}{et al.\xspace}

\newcommand{\revised}[1]{\textcolor{black}{#1}}
\usepackage{listings}
\usepackage{xparse}
\newsavebox{\fminipagebox}
\NewDocumentEnvironment{fminipage}{m O{\fboxsep}}
 {\par\kern#2\noindent\begin{lrbox}{\fminipagebox}
  \begin{minipage}{#1}\ignorespaces}
 {\end{minipage}\end{lrbox}%
  \makebox[#1]{%
    \kern\dimexpr-\fboxsep-\fboxrule\relax
    \fbox{\usebox{\fminipagebox}}%
    \kern\dimexpr-\fboxsep-\fboxrule\relax
  }\par\kern#2
 }

\begin{document}

\title{On the Adoption and Effects of Source Code Reuse on Defect Proneness and Maintenance Effort}

\author{
Giammaria Giordano,$^1$ Gerardo Festa,$^1$ Gemma Catolino,$^2$\\Fabio Palomba,$^1$ Filomena Ferrucci,$^1$ Carmine Gravino$^1$\\

\IEEEauthorblockA{$^1$Software Engineering (SeSa) Lab, Department of Computer Science - University of Salerno, Italy\\$^2$Jheronimus Academy of Data Science -- Tilburg University, 's-Hertogenbosch, Netherlands
\\
giagiordano@unisa.it, g.festa22@studenti.unisa.it, g.catolino@tilburguniversity.edu\\fpalomba@unisa.it, fferrucci@unisa.it, gravino@unisa.it
}
}

\maketitle
\thispagestyle{plain}
\pagestyle{plain}

\begin{abstract}
Context. Software reusability mechanisms, like inheritance and delegation in Object-Oriented programming, are widely recognized as key instruments of software design. These are used to reduce the risks of source code being affected by defects, other than to reduce the effort required to maintain and evolve source code. Previous work has traditionally employed source code reuse metrics for prediction purposes, e.g., in the context of defect prediction.  
%costs and effort required to build, maintain, and evolve software projects. In Object-Oriented programming languages, software reusability is supported through implementation inheritance, specification inheritance, and delegation. While previous work focused on the relationship between the adoption of these mechanisms and code quality, still little is known on how they impact the evolution of fault-proneness and code churn, i.e., two well-established proxy indicators of source code quality and effort.
%the number of lines of code necessary to fix bugs. %\textcolor{red}{CG: chiariamo subito che intendiamo per effort estimation?}.

Objective. However, our research identifies two noticeable limitations of current literature. First, still little is known on the extent to which developers actually employ code reuse mechanisms over time. Second, it is still unclear how these mechanisms may contribute to explain defect-proneness and maintenance effort during software evolution. We aim at bridging this gap of knowledge, as an improved understanding of these aspects might provide insights into the actual support provided by these mechanisms, e.g., by suggesting whether and how to use them for prediction purposes. 

Method. We propose an exploratory study aiming at (1) assessing how developers use inheritance and delegation during software evolution; and (2) statistically analyze the impact of inheritance and delegation on fault proneness and maintenance effort. The study will be conducted on the commits of 17 Java projects of the \textsc{Defects4J} dataset. 
\end{abstract}

\IEEEpeerreviewmaketitle

\section{Introduction}
\label{sec:introduction}
Software reusability is the design principle that allows developers to reuse part of the existing code to implement new features \cite{bieman1995reuse,soundarajan1998inheritance}. This practice is widely recognized as one of the key assets of software development, as developers may have multiple benefits, such as the reduction of evolution time, effort, and cost, other than of the risks of source code being affected by defects \cite{singh2010reusability,lange1989some,sharma2009reusability}.

When it turns to Object-Oriented programming languages, many software reuse mechanisms have been provided over time. Design patterns \cite{de2009design,gamma1993design}, third-party libraries \cite{zaimi2015empirical,salza2020third}, and programming abstractions \cite{sommerville2011software} are examples of these mechanisms. Focusing on \textsc{Java}, two very well-known types of programming abstractions are provided to developers: \emph{inheritance} and \emph{delegation} \cite{arnold2005java}. The former allows a class to take the properties and attributes of another class, establishing a hierarchical relation between them. The latter refers to when a class invokes an instance of another class to carry out operations without performing any other type of action.

The importance of these mechanisms has been remarked multiple times by researchers. Already in early 90s, Chidamber and Kemerer \cite{chidamber1994metrics} included the Depth of Inheritance Tree (DIT), i.e., a metric that measures the number of classes that inherit from another class, in their Object-Oriented metrics suite. Later on, several other metrics capturing various aspects of inheritance \cite{breesam2007metrics,mal2013new,rajnish2008class} and delegation \cite{cherkaoui1998qos,munro2005product,vanhilst2007reverse} were proposed, along with best and bad practices on how to use reusability mechanisms \cite{haefliger2008code,jalender2012designing,mantyla2003taxonomy,palomba2014mining}. On the empirical standpoint, a noticeable amount of investigations targeted the role of inheritance and delegation in keeping source code quality under control. For instance, researchers have been studying the relation between these mechanisms and other Object-Oriented metrics \cite{chhikara2011evaluating, chawla2013evaluating, e1996evaluating}, design patterns \cite{ampatzoglou2015effect, huston2001effects}, code complexity \cite{albalooshi2017comparative}, and source code maintainability \cite{daly1996evaluating,giordanoevolution,PRECHELT2003115}. 

Perhaps more interestingly, inheritance and delegation metrics have been often employed in the context of predictive models for software maintenance and evolution. The key example is defect prediction \cite{hall2011systematic,hosseini2017systematic}, where researchers assessed how reusability mechanisms may contribute to the prediction of future source code defects \cite{basili1996validation, singh2010empirical,yu2002predicting, di2017developer,palomba2017toward}. Similarly, the contribution of inheritance and delegation has been experimented for predicting maintenance effort change \cite{catolino2020improving,nagappan2005use}, code smells \cite{arcelli2016comparing,di2018detecting}, software vulnerabilities \cite{shin2010evaluating}, and infrastructure-as-code quality \cite{dalla2021within}. 

Despite the availability of a large body of knowledge on how inheritance and delegation mechanisms contribute to the prediction of source code attributes, most of the prediction models defined so far made a strong assumption: \emph{developers make use of reusability principles while evolving source code}. 

On the one hand, the extent to which these mechanisms are used in practice might have a notable impact on their contribution to prediction models. On the other hand, it is unclear how the relation between reusability and source code attributes varies over time and, therefore, whether inheritance and delegation mechanisms should still be considered for prediction purposes as the system evolves. 

In this registered report, we propose the methodology we plan to use to fill the limitations of current research with respect to the adoption of reusability practices and their evolutionary effects on two specific source code attributes such as \emph{defect proneness} and \emph{maintenance effort}. We select these attributes as they represent two interesting use cases to assess reusability mechanisms. First, these mechanisms are indeed supposed to reduce fault proneness and maintenance effort \cite{singh2010reusability,lange1989some,sharma2009reusability}. Second, a number of prediction models targeted the early location of defects and estimation of the effort required to perform evolutionary tasks \cite{catolino2020improving,pascarella2019fine,nagappan2005use}.

Our study will focus on \textsc{Java} projects, as Java (1) offers mechanisms that encourage the use of inheritance and delegation \cite{craig2007inheritance, tempero2013programmers} and (2) is still among the most popular programming languages used in industry.\footnote{Programming language ranking - Year 2021: \url{https://www.tiobe.com/tiobe-index/}} 
To conduct our experiment, we will first mine the \textsc{Defects4J} dataset to extract commit-level information on the adoption of reusability mechanisms. Then, we will develop statistical models to assess the contribution of reusability mechanisms on defect proneness---as indicated by the number of defects over time---and maintenance effort---as indicated by the code churn of commits. 

Our work has an \emph{exploratory} connotation, as we do not start with predefined hypotheses but plan to develop a set of hypotheses after the execution of the study, based on the results achieved. All the collected data and the scripts developed in the context of our research will be made publicly available for the research community.

\section{Background and Related work}
\label{sec:related}
We first describe the most widely used paradigms in the Object-Oriented programming languages for reusing code: \emph{inheritance} and \emph{delegation}. Then, we survey the related literature targeting code reusability and its impact on source code. 

\subsection{Inheritance and Delegation}
In \textsc{Java} there are two ways to define a hierarchical dependency between two classes:
\begin{description}[leftmargin=0.3cm]

\smallskip
  \item[\texttt{`extends'}.] Given two classes \texttt{A} and \texttt{B}, \texttt{A} is defined as super-class of \texttt{B} if \texttt{B} inherits variables or methods by \texttt{A}. In \textsc{Java} to establish  this super-class -- sub-class relation 
  the sub-class must indicate it through the keyword ``extends''.
  
  \smallskip
  \item[\texttt{`implements`}.] Given a class \texttt{A}, and an interface \texttt{B}, we will claim that \texttt{A} inherits from \texttt{B} if \texttt{A} implements the interface \texttt{B}. In \textsc{Java} this mechanism is provided using the keyword ``implements''. In particular, when a class \texttt{A} inherits using an interface, it must provide a concrete implementation of methods defined as a blueprint on interface.
\end{description}

These definitions recall the concept of \emph{reusability} in terms of specification inheritance, implementation inheritance, and delegation \cite{Brgge2009ObjectOrientedSE}.
From a practical point of view, the first one refers to the possibility of replacing an object \texttt{A} with an object \texttt{B} using a combination of two principles:

\begin{itemize}
     \smallskip
    \item \textbf{Strict Inheritance.} When a sub-class \texttt{A} exposes behavior and properties of super-class \texttt{B} without making any changes  \cite{Brgge2009ObjectOrientedSE}.
    
    \smallskip
    \item \textbf{The Liskov Substitution Principle.} According to Liskov and Wing \cite{liskov1994behavioral}, given two classes \texttt{A} and \texttt{B},  \texttt{A} is a sub-class of \texttt{B} if is possible to substitute the object \texttt{B} with the object \texttt{A} every time that the object \texttt{B} was expected.
\end{itemize}

The implementation inheritance occurs when a class  indirectly reuses a super-class source code. The sub-class can wholly or partially override methods and/or properties and replace the super-class's original behavior with its own. 
However, the implementation inheritance by definition violates the encapsulate principle because a sub-class could accidentally invoke methods or use some proprieties of the super-class in a wrong manner \cite{Brgge2009ObjectOrientedSE}. To avoid this, it is possible to substitute the implementation inheritance with the delegation in some cases. With this mechanism, a class \texttt{A} does not inherit anything from another class \texttt{B}, but \texttt{A} invokes methods of \texttt{B} directly by declaring itself a variable of type B.

\subsection{Related Work}
%In the recent past, many researchers have explored multiple ways code reusability, effort estimation, and fault-proneness from different perspectives and outcomes. We discuss each subtopic individually to summarize better the state of the arts and limitations of these works and their difference. 
%We started from the code reusability side; afterward, we moved on to the fault-proneness, and finally, we will focus on the effort estimation.
Source code reusability has been the subject of several researches in the last decades. These touched various angles of the problem, by introducing novel metrics to capture inheritance relations \cite{chidamber1994metrics, breesam2007metrics,mal2013new,rajnish2008class} and delegation \cite{cherkaoui1998qos,munro2005product,vanhilst2007reverse}, defining best design practices to exploit the benefits of reusability \cite{haefliger2008code,jalender2012designing}, or identifying a number of source code quality issues that reusability can cause, e.g., code smells \cite{mantyla2003taxonomy,palomba2014mining,fowler2018refactoring}. While the scope of our work targets inheritance and delegation mechanisms, it is worth mentioning the existence of close research areas such as the analysis of design patterns \cite{fontana2013design,zhang2013survey} and third-party libraries \cite{zhan2021research}. 

\smallskip
\textbf{Reusability and code quality.} As for the themes of our exploratory study, Albalooshi and Mahmood \cite{albalooshi2017comparative} conducted an empirical analysis on the implementation inheritance by considering three programming languages like \textsc{C++}, \textsc{Python}, and \textsc{Java}. %As a result, the authors found that the mechanisms of \textsc{Java} to define inheritance mechanisms tend to degrade the quality of source code. Goel and Bathia \cite{goel2013analysis} obtained similar results by conducting an analysis on the impact of multilevel inheritance on the reusability considering three \textsc{C++} projects. 
 As a result, the authors found that the mechanisms of \textsc{Java} to define inheritance tend to degrade source code quality. Goel and Bathia \cite{goel2013analysis} obtained similar results by analyzing the impact of multilevel inheritance on the reusability considering three \textsc{C++} projects. They found a negative correlation between the use of inheritance and the quality of source code in terms of maintainability. Other research efforts targeted the effect of inheritance and delegation on various aspects of source code quality. Chhikara et al. \cite{chhikara2011evaluating} conducted a case study on one small-scale software project, reporting on the correlation between inheritance metrics and other metrics belonging to the Chidamber and Kemerer suite. Chawla and Nath \cite{chawla2013evaluating} took a closer look at how inheritance and delegation metrics may impact software coupling, concluding that these metrics can be useful to assess code quality. Similar findings were reported by Abreu et al. \cite{e1996evaluating}. Additional experiments were conducted to assess the relation between reusability and design patterns \cite{ampatzoglou2015effect, huston2001effects} and code complexity \cite{albalooshi2017comparative}: all these studies converged toward the relevance of inheritance and delegation. More recently, we carried out a study to investigate the evolution of inheritance and delegation and their impact on the severity of code smells \cite{giordanoevolution}. The results revealed that inheritance and delegation tend to increase over time, but not in a statistically significant manner. However, increasing the adoption of these mechanisms tends to decrease code smells' severity. 

The potential benefits of reusability have led researchers to use inheritance and delegation metrics within prediction models. In this respect, most of the defect prediction models include reusability as a feature \cite{hall2011systematic}. Perhaps more importantly, these metrics have been sometimes shown to significantly contribute to the predictions of those models: for instance, Jureczko and Madeyski \cite{jureczko2010towards} showed that the Depth of Inheritance Tree metric is among the best predictors of source code defectiveness. These results were later confirmed by other software maintenance and evolution researches \cite{singh2017software,jureczko2010using}. 

\smallskip
\textbf{Reusability and maintenance effort.} From an empirical side, Prechelt \etal \cite{PRECHELT2003115} carried out two experiments to investigate the relation between inheritance metrics and maintenance effort estimation. Their results revealed that maintaining a low level of inheritance depth positively impacts the (decrease of) developer's effort to maintain source code. Similarly, Daly \etal \cite{daly1996evaluating} showed that as the inheritance depth level increases, so does the effort of developers to maintain code. 

In terms of maintenance effort estimation, researchers have been mainly looking at process-level information (e.g., team data and measurements of the development activities), attempting to provide indications in terms of direct and indirect estimations of entire projects under maintenance \cite{wu2016maintenance}. Besides that, researchers have been also working on effort prediction of maintenance activities, which revolves around the prediction of the effort spent in performing specific activities such as code review \cite{mishra2014mining} and bug fixing time \cite{anbalagan2009predicting,bougie2010comparative}. The contribution provided by reusability metrics to those models are, however, unclear. Recently, Nagappan et al. \cite{nagappan2005use} and Liu et al. \cite{liu2017code} proposed the use of code churn, i.e., the amount of lines of code modified within commits, as an alternative metric of maintenance effort which better aligns with the actual effort spent by developers while performing evolutionary tasks.

\smallskip
\textbf{Our work.} With respect to the papers discussed above, ours has multiple differences. In the first place, most of previous work analyzed reusability by relying on the computation of metrics, e.g., DIT; as further elaborated in Section \ref{sec:design}, we plan to operationalize reusability by means of specification inheritance, implementation inheritance, and delegation, being able to better map the employment of reuse mechanisms over time. In the second place, we plan to conduct a fine-grained analysis where the evolution and impact of reusability will be investigated at commit-level. Furthermore, we plan to address a key limitation of most previous work proposing prediction models: the contribution of code reuse to their capabilities indeed assumes that developers make use of reusability mechanisms. As such, our evolutionary study will provide more detailed insights into the potential benefits brought by inheritance and delegation to state-of-the-art prediction models.

\begin{figure*}
    \centering    
    \includegraphics[width=1\textwidth]{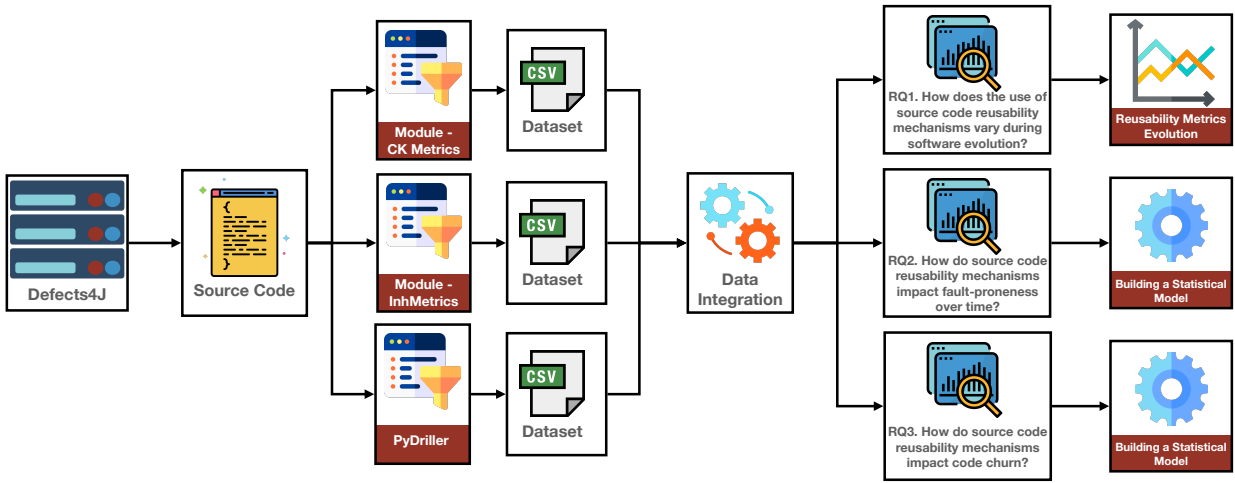}
    \caption{Overview of the methodology applied to address our research questions.}
    \label{fig:overview}
\end{figure*}

\section{RESEARCH QUESTIONS AND OBJECTIVES}
\label{sec:ResearchQuestions}
%\textcolor{red}{Carmine: hp provato a dare goal con il solito template -}

The \emph{goal} of the study aims at investigating how the use of reusability mechanisms evolves over time and assessing their impact on fault-proneness and code churn. The \emph{purpose} is to understand whether those mechanisms can provide developers with an indication of source code quality variation---considering the fault-proneness and effort to fix faults of a project. The \emph{quality focus} is on the reusability in terms of implementation inheritance, specification inheritance, and delegation and their evolution within software projects. 
We conduct the analysis from both practitioners and researchers (\emph{perspective}). The first one is interested in understanding whether the reusability mechanism can be suitable for monitoring the quality of a system. The latter is to have more evidence about inheritance and delegation mechanisms when monitoring the source code quality. The \emph{context} of our investigation will be JAVA projects publicly available.
Based on the goal of our study, we formulate three main research questions.
\begin{center}
	\begin{rqbox}
		%\textbf{RQ1.} \emph{How do the use of reasubility mechanisms evolve over time?}
		\textbf{RQ1.} \emph{How does the use of source code reusability mechanisms vary during software evolution?}
	\end{rqbox}
\end{center} 
\revised{The first research question aims at understanding the use of source code reusability mechanisms by developers during software evolution. More specifically, the goal of \textbf{RQ1} is that of providing insights on the evolution of reuse mechanisms that might later be exploited to better interpret the findings of \textbf{RQ2} and \textbf{RQ3}. In other terms, the patterns observed in the context of this research question will be also useful to understand the effects of inheritance and delegation on defect-proneness and code churn, e.g., should we identify an exponential growth in the adoption of delegation, this would potentially make this mechanism more relevant for software evolution, hence influencing more the amount of code churn required to apply modifications.}
%\gemma{non creda sia come l'adottano, perche noi in realtà vediamo solo l'evoluzione in termini di quantità e non qualità..o sbaglio? (carmine: condivido - modifica pure}
%\giammaria{L'ho cambiata...va meglio?}
Since we intend to analyze three mechanisms for \textsc{reusability}, \ie specification inheritance, implementation inheritance, and delegation \cite{Brgge2009ObjectOrientedSE}, that can impact differently on software evolution, we will consider three sub-research questions:

\smallskip
\begin{description}[leftmargin=0.9cm]
    %\item[RQ1$_1$.] \emph{How does the use of the implementation inheritance vary during software evolution?}
    \item[RQ1$_1$.] \emph{How does the use of implementation inheritance vary during software evolution?}
    \smallskip
    \item[RQ1$_2$.] \emph{How  does the use of the specification inheritance vary during software evolution?}
    
    \smallskip
    \item[RQ1$_3$.] \emph{How does the use of delegation vary during software evolution?}
\end{description}

%\textcolor{red}{metriche ancora non introdotte}    
Once the evolution of reusability mechanisms is analyzed, we plan to investigate how the evolution might affect code quality, measured in terms of fault-proneness. 
%Once analyzing how these metrics evolve, we focus on understanding if exist a correlation between the code quality metrics considered and the presence of bugs in source code. Hence, we asked:
%The second research focuses on understanding whether  

\begin{center}
	\begin{rqbox}
		\textbf{RQ2.} \emph{
		How do source code reusability mechanisms impact fault-proneness over time?
		}
	\end{rqbox}
\end{center} 
\smallskip
\revised{Finally, we plan to assess the impact of reusability mechanisms on the maintenance effort required to fix faults. Among the various \emph{direct} and \emph{indirect} metrics available in literature \cite{wu2016maintenance}, we will operationalize maintenance effort through \emph{code churn}, that is, the amount of lines of code modified within a commit. This is an indirect metric that can proxy the actual effort spent by developers when maintaining source code \cite{wu2016maintenance,munson1998code,mcintosh2011empirical}.}
%Finally, we want investigate on the impact of quality metrics on the effort estimation problem. For this reason, we formulated the following research question
\begin{center}
	\begin{rqbox}
		\textbf{RQ3.} \emph{How do source code reusability mechanisms impact code churn?}
	\end{rqbox}
\end{center}

The above research questions will be addressed by employing statistical tests and models (see details in Section \ref{sec:experimental}). 
% we will approach each research question individually using different approach based on the specific \textbf{RQ} that we want to respond.
To design and report on the  empirical study to be performed, we will follow the guidelines proposed by Wohlin  \etal\cite{wohlin2012experimentation} and \emph{ACM/SIGSOFT Empirical Standards}\footnote{Available at: \url{https://github.com/acmsigsoft/EmpiricalStandards}}. All the experimental material (e.g., datasets, scripts) will be publicly available in an online appendix.

%To conduct our investigation, we will approach each research question individually using different 
%To respond to our first research question, we will analyze the behavior of inheritance and delegation over time for each project considered

%statistically analyze the impact of inheritance and delegation on fault-proneness and code churn
%\input{reusability_evolution_impact/Sections/design}
\begin{table*}[h]
	%\tiny
	\centering

	\resizebox{1\linewidth}{!}{
		\begin{tabular}{|l|r|r|r||r|r|r|r|r|r|r|}
				\hline
			\rowcolor{arsenic} {\textcolor{white}{\textbf{Project Name}}}&
			{\textcolor{white}{\textbf{Number of Bugs}}} & 
			{\textcolor{white}{\textbf{Active Bug Ids}}} &
			{\textcolor{white}{\textbf{Deprecated bug Ids}}}& 
			{\textcolor{white}{\textbf{Pull Request}}} &
			{\textcolor{white}{\textbf{Contributors}}} &
			{\textcolor{white}{\textbf{Stars}}} &
			{\textcolor{white}{\textbf{Forks}}} &
			{\textcolor{white}{\textbf{Commits}}} &
			{\textcolor{white}{\textbf{Branches}}} &
			{\textcolor{white}{\textbf{LOC}}} 

						\\

\rowcolor{gray10}JFreeChart	& 26 & 1-26 & None & 22 & 24 & 866 & 355 & 4218	& 3 & %291898 (la vera repo non è su github)
~250k - ~290k\\
Commons-Cli & 39 & 1-5,7-40 & 6 & 8 & 42 & 255 & 154 & 1169 & 4 & %16843
~5k - ~16k \\
\rowcolor{gray10} Closure-Compiler & 174 & 1-62,64-92,94-176 & 63,93 & 6 & 472	& 6,5k & 1,1k & 17,962 & 76 & %1218641
~60k - ~60k\\
Commons-Codec & 18	& 1-18	& None	& 9	& 40 & 364 &207 & 2,244	& 7 & %55432
~48k - ~34k\\
\rowcolor{gray10} Commons-Collections & 4 &25-28 &1-24 & 37 & 62 & 551	& 389	& 3,729 & 8 & %136982
~49k - ~60k\\
Commons-Compress & 47 & 1-47 & None & 9 & 67 & 231 & 210 & 3,602 & 9 & %362366
~129k - ~91k\\
\rowcolor{gray10} Commons-Csv & 16 & 1-16 & None & 8 & 37 & 281 & 220 & 1,796 & 4 &	%170278
~166k - ~166k\\
Gson & 18 & 1-18 & None & 151 & 125 & 21,2k	& 4,1k & 1,668 & 14 & %43422
~68k - ~70k\\
\rowcolor{gray10} Jackson-Core & 26	& 1-26 & None & 2 & 63 & 2.1k & 690	& 2,124 &	21 & %780517
33k - ~66k\\
Jackson-Databind & 112 & 1-112	& None & 19 & 198 & 3,1k & 1,2k	& 6,578	& 22 &	%106462
~98k - ~235k\\
\rowcolor{gray10} Jackson-Dataformat-XML & 6 & 1-6 & None & 3 & 26 & 497 & 189 & 1,318 & 19 & %277883
~59k - ~117k\\
JSoup & 93 & 1-93 & None & 43 & 99 & 9,6k & 2k & 1,693 & 3 & %40816
~39k - ~34k\\
\rowcolor{gray10} Commons-JXPath	& 22	& 1-22	& None	& 8	& 17 & 18 & 40 &	601	& 4	& %47040
~46k - ~26k\\
Commons-Lang & 64 & 1,3-65 & 2 & 92 & 174 & 2,3k & 176 & 6,859 & 8 & %189720
~160k - ~190\\
\rowcolor{gray10} Commons-Math	& 106	& 1-106	& None & 68	& 48 & 451 	& 71		& 7,004	& 17 & %7401
~58k - ~63k\\
Mockito & 38 & 1-38 & None & 7 & 246 & 13,1k	& 2,3k & 5,787 & 16 & %109717
~73k - ~94k\\
\rowcolor{gray10}Joda-Time & 26 & 1-20,22-27 & 21 & 2 & 77 & 4,8k & 922 & 2,196 & 6 & %178857
~103k - ~164k\\
			\hline
		\end{tabular}}
	\caption{\revised{Characteristics of the projects considered in the study. Information concerned with the \textsl{`Active Bug IDs'} and \textsl{`LOC'} are provided in a range reporting the minimum and maximum values observed over the history of the projects.}}
	\label{table:defects4j_dataset}
\end{table*}

\section{Research Methodology}
\label{sec:design}

%\textcolor{red}{Carmine: secondo me andrebbe organizzato diversamente. c'è data extraction e poi datasets (e c'è sovrapposizione tra cappello della sezione e l'inizio della sezione dataset). Poi experimtal plan presentato in una sezione a parte. Farei le seguenti sottosezioni: Dataset, Data extraction procedure, Experimental plan}

Figure \ref{fig:overview} overviews the methodology we intend to exploit to address our research questions, as detailed in the next sections.

%To answer to our research questions, we plan to perform an empirical analysis on \textsc{Java} projects provided by \textit{Defects4J} (detailes in  Section  Section \ref{sec:dataset}). 

%According to our preliminary analysis, we estimate that we will analyze over 9,000 commits.
%Due to the lack of investigation of the evolution of inheritance and delegation and their impact on code quality in commit level grain, we will focus on filling the gap and performing our work considering the evolution in terms of new commits.
  
%In the following we describe the experiment procedure that we will use to address our research questions.

\subsection{Dataset}
\label{sec:dataset}
We plan to perform an empirical analysis on \textsc{Java} projects provided by \textsc{Defects4J}, which collects information on 835 bugs provided by 17 real \textsc{Java} projects. 
According to the official documentation\footnote{https://github.com/rjust/defects4j} each bug collected into the dataset is characterized by the following properties: 
\begin{enumerate}
    \item It is fixed in a single commit, meaning that the bug resolution never refers to more than one commit;
    \item It is minimized, meaning that Defects4J maintainers manually remove commits that do not provide information about the introduction of bugs or fixing activity (e.g., commits where refactoring activities are done);
    \item The fixing activities modify the source code. This means that the bug introduction can be caused by several factors, \eg wrong parameters in configuration files and problems in the production class. However, the corresponding fixing only concerns changing the source code.
\end{enumerate}

\revised{There are multiple reasons leading to the selection of this dataset. First, it enables the investigation of the impact of reuse mechanisms in a \emph{noise-free} environment. Indeed, we can provide more precise insights into the actual role played by inheritance and delegation which would not be possible through larger software repository mining studies where the existence of uncontrolled conditions, e.g., tangled changes \cite{herzig2016impact}, may bias the conclusions provided. Secondly, despite the defects being carefully selected, those defects are of different types and nature, therefore representing various defects affecting real-world software systems \cite{sobreira2018dissection}. Last but not least, Defects4J has been widely used in literature (e.g., \cite{martinez2017automatic,durieux2015automatic}), hence representing a valuable asset that enables us to build additional knowledge on a state-of-the-art dataset - this would also be useful for other researchers interested in building on top of our work.}

As mentioned in Section \ref{sec:related}, little has been done to analyze code reuse mechanisms over time and how those may contribute to explaining fault-proneness and maintenance efforts during software evolution. For this reason, we intend to fill the gap by analyzing code reuse mechanisms from a low granularity perspective, \ie commits. We plan to analyze over 9,000 commits. \revised{Table \ref{table:defects4j_dataset} reports statistics of the projects included in  the Defects4J dataset. In particular, for each project the table provides (i) the numbers of defects, (ii) the IDs of fixed and unfixed defects, (iii) process metrics such as numbers of commits, numbers of pull request, and number of contributors; and (iv) its minimum and maximum LOC.}
% As mentioned in Section \ref{sec:related}, little has been done to quantify and analyze the use code reuse mechanisms over time by developers. Similarly, it is still unclear how these mechanisms may contribute to explaining fault-proneness and maintenance efforts during software evolution. \gemma{citare studi}.
% \gemma{qui sopra pero non parla di release come arriviamo a parlare di commit?Ci manca qualcosa qui?}

% \textcolor{red}{carmine: da quello scritto alla fine del related work, non ci dovrebbero essere studi che investigano reusability mechanisms at commit level. giusto? quindi forse conviene mettere una frase che dice che lavoriamo a livello di commit e poi diciamo che prevediaom di usare 9,000 commits }

% Thus, we want to fill this gap by considering the evolution in terms of new commits in our empirical analysis. 
% According to our preliminary analysis looking at the number of projects on GitHub, we estimate to analyze over 9,000 commits.

%\gemma{quale?}, we estimate that we will analyze over 9,000 commits.

%Due to the lack of investigation of the evolution of inheritance and delegation mechanisms and their impact on code quality in commit level grain, we will focus on filling the gap and performing our work considering the evolution in terms of new commits.

\subsection{Data Extraction Procedure}
To answer our research questions, we need to quantify the reusability in terms of implementation inheritance, specification inheritance, and delegation.
To this end, we plan to use a tool already validated in our previous work \cite{giordanoevolution}. In particular, the tool computes those metrics following these patterns:

%First step of our study will perform with the reusability metrics, focusing on inheritance and delegation. To extract and elaborate these metrics, we will use a self-developed tool also validate during our previous work \cite{giordanoevolution}. To calculate these metrics, the tool computes the following logical operations:

\smallskip
\begin{description}[leftmargin=0.3cm]
    \item[Specification Inheritance.]
    Given a class \textit{A}, the tool considers the specification inheritance as the arithmetical sum of each interface used by \textit{A}. For instance, suppose that \textit{A} inherits methods from two interfaces \textit{B} and \textit{C}, and \textit{C} in turn inherits methods from another interface \textit{D}. In this case, the specification inheritance for \textit{A} is 3.
        
    \smallskip
    \item[Implementation Inheritance.]
    Suppose that \textit{A} is a sub-class of \textit{B}, the tool considers the implementation inheritance
    as the arithmetical sum of each method in \textit{B} called by some method in \textit{A}. For example, suppose that \textit{A} is a class with N methods, and \textit{B} a class with just one method call \texttt{bar()}. To increase the number of implementation inheritance by one, one of the methods in \textit{A} must invoke \texttt{bar()}. %It is necessary to note that if \textit{B} in turn is a sub-class of another class \textit{C}, also the methods in \textit{C} will consider to the computation.
    %it is important to note that this sum increases as the entire inheritance tree is navigated, e.g., if \textit{B} is a sub-class of another class \textit{C},containing 5 methods than the sum will be.
    
    \smallskip
    \item[Delegation.]
    %Given a class \textit{A}, the tool considers the delegation metric as the arithmetical sum of each non-primitive variable (e.g., int, double, String and so on) or have a \textit{non-binding} type (e.g.,  Checkbox variables provides by and external library). For each of them, the tool verifies if this variable is used only to invoke external objects.
    Given a class \textit{A}, the tool considers the delegation metric as the arithmetical sum of each non-primitive variable (i.e., variables different from \texttt{int}, \texttt{double}, \texttt{String}, and so on) or variables that do not have a binding type provided by external libraries (e.g., \texttt{Checkbox} offered by \texttt{javax.swing} framework). For each variable, the tool verifies if it is only used to invoke external objects.
\end{description}

\smallskip
To answer research questions \textbf{RQ2} and \textbf{RQ3}, we plan to collect information on bugs and code churns. \textit{Defects4J} contains information on bugs at commit level, while we will consider 
\textsc{PyDriller}, an automatic static analysis tool that can analyze \textsc{Git} repositories, to extract information about commits, developers, modifications, diffs, and
source code \footnote{https://pydriller.readthedocs.io/en/latest/intro.html}.

%from a methodology standpoint, we will combine the information provided by our tool with \textsc{PyDriller} an automatic static analysis tool that can analyze repositories under the version control \textsc{Git} and can be use to extract information as about commits, code churns and so on.\footnote{https://pydriller.readthedocs.io/en/latest/intro.html}.
%We want combine the information provided by PyDriller with the information extracted by our tool, and we want apply statistical tests to address the remaining \textbf{RQs}.
%\input{reusability_evolution_impact/Sections/experimental_plan}
\subsection{Experimental Plan}
\label{sec:experimental}
%The following section describe the approach we will use to respond to our research questions. For understandability reasons, we discuss each research question individually.

For the sake of comprehensibility, we present the empirical analysis we plan to perform for each research question.

\begin{enumerate}
    \item \emph{RQ1. Analysis of the evolution of reusability mechanisms  over time.} To answer this research question we will analyze the 
    behavior of the reusability metrics (implementation inheritance, specification inheritance and delegation) during the evolution. In particular, we plan to employ basic statistical analysis, and visualize results using plots.
    
    \smallskip
    \item \emph{RQ2. Analysis of the impact on fault-proneness of reusability mechanisms over time.} Moving on  \textit{RQ2}, we plan to build a statistical model to verify how reusability metrics impact the variability of bugs in the source code.
    
    \smallskip
    \item \emph{RQ3. Analysis of the impact on maintenance effort of reusability mechanisms over time.} Also for \textit{RQ3}, we plan to build a statistical model to verify how reusability metrics impact the maintenance effort to fix a bug.
\end{enumerate}
%\subsubsection{RQ1. Analysis of the evolution of reusability mechanisms  over time}
%To answer this research question we will analyze the 
%behavior of the reusability metrics (implementation inheritance, specification inheritance and delegation) during the evolution. In particular, we plan to employ basic statistical analysis, and visualize results using plots.

%we want analyze the reusability metrics distributions - implementation inheritance, specification inheritance and delegation, in order to understand the behaviours during the evolution. 
% To increase the reliability, we will normalize the sample using the LOC to focus on the relative frequency rather than the absolute frequency.
% Indeed, it is reasonable to assume that during the natural evolution of software, developers tend to use inheritance and delegation mechanisms, so their absolute value increase would be a trivial result.
%For these reason, we will focus on the relative frequency rather than the absolute frequency. 

%\subsubsection{RQ2. Analysis of the impact
%on fault-proneness of reusability mechanisms
% over time}
%correlation between the code quality metrics and the presence of bugs in source code
%Moving on  \textit{RQ2}, we plan to build a statistical model to verify how reusability metrics impact the variability of bugs in the source code. 

%\subsubsection{RQ3. Analysis of the impact
%on maintenance effort of reusability mechanisms
% over time}
%As for \textit{RQ3}, we plan to build a statistical model to verify how reusability metrics impact the maintenance effort to fix a bug. 

The statistical models will be devised as follows.

\begin{description}[leftmargin=0.3cm]
\smallskip
\item[Independent Variables.]
According to our previous considerations, we will use as independent variables the reusability metrics, \ie implementation inheritance, specification inheritance, and delegation.

\smallskip
\item[Response Variable.]
The number of bugs represents our response variable. However, since our goal is to understand the variability of them, we plan analyzing whether the number of bugs between two commits is \textit{stable}, or  \textit{increase}/\textit{decrease}.
In particular, it will be considered  ``stable'' if we do not identify any changes in terms of the number of bugs between the commit \textit{I} and the commit \textit{I+1}. It will be considered as  ``increase'' (``decrease'') if we identify a positive (negative) value as a result of the subtraction between the numbers of bugs on the commit \textit{I+1} and the numbers of bugs on the commit \textit{I}. 
%\gemma{forse la parte di giu potrebbe essere messa nelle threats?}

% -- it means that a bug could be introduced in a \textsc{Java} file, an \textsc{XML} file, and so on. However, the respective fix  has always done modifying one or more \textsc{Java} files. Considering this observation, we will filter out bugs introduced  by \textsc{XML} files, configuration files, and so on.
%To avoid possible threats to validity we want discard of our analysis 
%To avoid possible threats to validity since considering them could go into analyzing bugs that were fixed by changing a certain \textsc{Java} file but were not introduced by any \textsc{Java} file and this could lead to inaccurate analysis.  
 
\smallskip
\item[Control Variables.]
Conscious that bugs variation could depend on other external factors, we will consider two  sets of metrics as control variables. On the one hand, we will consider the following Chidamber and Kemerer (CK) metrics \cite{chidamber1994metrics}: \emph{DIT} (Depth of Inheritance Tree), \emph{NOC (Number Of Children)}, \emph{LOC} (Lines of Code), \emph{LCOM} (Lack of Cohesion of Methods), \emph{WMC} (Weighted Methods per Class), \emph{RFC} (Response for a Class) and \emph{CBO} (Coupling Between Objects).
On the other hand, we will also take into account the \emph{Code Churns}.
It is important to mark that although \emph{NOC} and \emph{DIT} are also metrics related to code reuse, we will include 
them with the intent of comparing their statistical power to the adoption mechanisms estimated by our reusability metrics.
However, we plan to assess the presence of possible multi-collinearity when performing the statistical modeling due to the presence of those related metrics. We will rely on previous guidelines provided by the literature \cite{allison2012can,lieberman2014precise}.

\smallskip
\item[Choosing Statistical Model.]
To address \textbf{RQ2}, we will use a \emph{Multinomial Log-Linear Model} \cite{theil1969multinomial}.
This model generalizes logistic regression to multi-class problems, so it perfectly fits our case. In particular, as already done in our previous work \cite{giordanoevolution}, we plan to use \textsc{R} for running the analysis using the function \textsc{multinom} available in the package \textsc{nnet}\footnote{https://cran.r-project.org/web/packages/nnet/nnet.pdf}---that fits the model via neural networks. Finally, as for \textbf{RQ3}, given the nature of the response variable, \ie code churn, we will use a different statistical model, \ie \emph{Generalized Linear Model} \cite{faraway2016extending} using \textsc{glm} function.
\end{description}

\revised{\textbf{Additional analysis.} The influence of the reuse metrics on introducing defects might not necessarily be directly measurable. For instance, previous work \cite{feigenspan2011exploring,katzmarski2012program,mihancea2009discovering} reported that inheritance might negatively impact program comprehension, which, in turn, can negatively contribute to the defect-proneness of source code. In other words, the value of reuse metrics can be sneakier, representing a co-occurring phenomenon rather than directly responsible for introducing defects. For this reason, besides interpreting the statistical codes provided by the statistical models, we will also (1) compute the number of cases in which defect-inducing commits involved the variation of inheritance and delegation metrics and (2) manually analyze those cases to better understand the way these metrics can directly impact the introduction of defects. Such an additional analysis will therefore provide more qualitative insights into how reuse mechanisms can impact defect-proneness.}

%\textcolor{red}{carmine: inseriamo un ref come abbiamo fatto per quella multinomial?}
% Indeed, differently from \textbf{RQ2}, the dependent variable  is a continuous variable, and as a consequence, it is possible to apply a prediction model that loose the constraints of the \emph{Multinomial Log-Linear Model}.

% \smallskip
% \giammaria{In the context of our study, we want to use Allison's guidelines\cite{allison2012can}, which outline how to manage a model for \emph{multi-collinearity} and when to ignore it.}

% \textcolor{red}{carmine: perchè qui citiamo un riferimento per gestire la multicollinearità mentre per RQ2 non lo abbiamo fatto?}
% \giammaria{convine forse dire nelle threats che abbiamo considerato sia per RQ2 che per RQ3 la possibilità di multi-collinearità? (ho verificato e festa l'ha considerata sia per la RQ2 che per la RQ3)}

%A fundamental aspect that we will consider to avoid some threats to validity is the \emph{multi-collinearity}. 
%To mitigate possible biasing of interpretation, we will follow the guidelines proposed by Allison \cite{allison2012can}
%In particular, we will verify a possible relationship among the independent variables using the \textsc{Kendall} correlation matrix. We will consider a correlation between two variables if the achieved correlation statistics is greater than 0.75. In case of correlation, we will discard the variable with a lower correlation with the dependent variable.

\subsection{Publication of generated data}
All the material of our study, \eg scripts will be publicly available in an online repository (\eg GitHub) to guarantee the replicability of our work and possible reuse for future investigations by other researchers.

\section{Threats To Validity}
\label{threats}
In this subsection, we discuss possible threats to validity and the strategies we will adopt to mitigate them.

\smallskip
\noindent \textbf{Construct Validity.}
These threats refer to a possible mismatch between the theory and the observation. \revised{Therefore, the selection of the dataset represents a crucial point. We plan to use Defects4J, which has been already widely used by the research community in several studies (e.g., \cite{DBLP:conf/wcre/SobreiraDDMM18}\cite{DBLP:journals/chinaf/JiangXX19} \cite{9286062}) and that will reduce possible bias due to the presence of uncontrolled conditions, \eg tangled changes \cite{herzig2016impact}, allowing us to investigate the impact of reuse mechanisms on defect-proneness and maintenance effort more precisely.}
\revised{A second threat to validity relates to the selection of the metric used to operationalize maintenance effort. We plan to use code churn \cite{munson1998code}: we are aware that this metric can only proxy the actual effort spent when maintaining source code, yet this choice is required in our case because of the unavailability of precise data regarding the maintenance effort in our dataset.} The tool we plan to use to extract metrics, \eg reusability or CK metrics, represents another potential threat to validity. We will use tools already validated and used by the research community \cite{giordanoevolution,spadini2018pydriller}.
Finally, as mentioned in Section \ref{sec:dataset}, in Defects4j a single bug can be introduced by multiple factors, but its resolution will always occur within a \textsc{Java} file. Thus, to avoid possible threats to contraction validity, we will discard commits that introduced bugs caused by issues not involving source code. This allows focusing only on defects introduced and resolved through changes to the source files.
%And lastly, the impact on branching during the mining process. This threat refers to an intrinsic characteristic of \textsc{GitHub}. Indeed, the platform manages in a tree structure the branching, which means that in some cases, the chronological order could suffer inconsistencies due to some merge or other operations (e.g., deletion of a branch). Researchers widely cover this threat, and according to Kovalenko \etal \cite{Kovalenko}, the inclusion of these branches does not statistically impact the defect prediction.

\smallskip
\noindent \textbf{Internal Validity.}
These threats refer to factors that can impact the study results. In our context, the threat concerns the metrics we intend to exploit to build the statistical models. Besides our hypothesis, we will use control variables ---previously shown to be  significant for source code quality \cite{tamburri2020success,succi2005empirical,chhikara2011evaluating,daly1996evaluating}---thus guaranteeing the reliability of our results.

\smallskip
\noindent \textbf{Conclusion Validity.} Threats related to this area refer to the selection and the use of the statistical test. In particular, for addressing \textbf{RQ2} we will use the Multinomial Logistic Linear Model \cite{theil1969multinomial}.
As for \textbf{RQ3}, we will apply the Generalized Linear Model \cite{faraway2016extending}.
These choices come from the nature of our response variables, \ie multiclass and continuous, respectively. Moreover, the research community used this type of model in similar contexts \cite{giordanoevolution, gemmaUnderstanding, lambiase2022_cultural_dispersion_community_smells}.%\gemma{citare paper gemma sulle variability community smell 2021 icse seis, paper nostro saner, paper stefano icse 2022}

\smallskip
\noindent \textbf{External validity.}
\revised{Threats in this category concern the generalizability of the results. Our work employs statistical analysis to seek relations between the employment of reuse mechanisms and source code maintainability, operationalized with defect-proneness and code churn metrics. The target of the work will be composed of 17 \textsc{Java} projects with over 9,000 commits coming from the \textsc{Defects4J} dataset. As such, our work is based on the analyses conducted on a \emph{sample}, hence our generalization strategy can be identified within the \emph{sample-based generalization} strategies proposed by Wieringa and Daneva \cite{wieringa2015six}. In particular, among those strategies, the ``statistical learning'' seems to be the most appropriate. Wieringa and Daneva \cite{wieringa2015six} reported that the \emph{``descriptions of statistical sample phenomena can be used to predict similar phenomena in new samples. [...]. The goal is not to generalize to a population, but to generalize to the next few cases''}. This strategy is basically in line with the \emph{generalizing by similarity} principle described by Ghaisas \etal \cite{ghaisas2013generalizing}. When contextualizing those strategies in our case, it is likely that similar results might be obtained in projects having similar characteristics with respect to those analyzed in our work (see Table \ref{table:defects4j_dataset}). Therefore, we cannot claim the generalizability of our findings to projects having different properties or even written in different programming languages. Replications in these contexts would still be desirable.}
\section{Conclusion}
\label{sec:conclusion}
This research will focus on understanding how inheritance and delegation mechanisms evolve over time and their impact on code quality, \eg variability of bugs, at the commit level. In particular, we will conduct this study on over 9,000 commits provided by 17 Java projects reclaimed from Defects4J. We first plan to analyze the evolution of reusability metrics at the commit level. Then we will construct two different statistical models for assessing whether reusability metrics---combined with additional factors---impact bugs variability and the time to fix bugs in terms of code churn.
%\gemma{non metterei future agenda anche perche questo non è un vero paper, nel senso la future è che facciamo il paper}.

% using our self-developed tool to extract and calculate the inheritance and delegation metrics for each commit. In addition, we want to combine this information with CK metrics provided by PyDriller in order to understand if these metrics can be used as a good predictor to estimate the effort necessary to fix bugs.

% As a future part of our agenda, we plan to expand our analysis, including additional projects, in order to confirm or refute the results obtained. 

\section*{Acknowledgment}
%Omitted for double-blind review.
Gemma is partially supported by the European Commission grant no. 825040 (RADON). Fabio is supported by the Swiss National Science Foundation through the SNF Project No. PZ00P2 186090 (TED). 

\balance
\bibliographystyle{IEEEtran}
\bibliography{reusability_evolution_impact/main}

\end{document}